# Core-Core Dynamics in Spin Vortex Pairs

S. S. Cherepov, B. C. Koop, V. Korenivski
*Royal Institute of Technology, 10691 Stockholm, Sweden*

D. C. Worledge
*IBM Watson Research Center, Yorktown Heights, NY 10598, USA*

A. Yu. Galkin, R. S. Khymyn, B. A. Ivanov
*Institute of Magnetism, Ukrainian Academy of Sciences, Kiev, Ukraine*

**We investigate magnetic nano-pillars, in which two thin ferromagnetic nanoparticles are separated by a nanometer thin nonmagnetic spacer and can be set into stable spin vortex-pair configurations. The 16 ground states of the vortex-pair system are characterized by parallel or antiparallel chirality and parallel or antiparallel core-core alignment. We detect and differentiate these individual vortex-pair states experimentally and analyze their dynamics analytically and numerically. Of particular interest is the limit of strong core-core coupling, which we find can dominate the spin dynamics in the system. We observe that the 0.2 GHz gyrational resonance modes of the individual vortices are replaced with 2-6 GHz range collective rotational and vibrational core-core resonances in the configurations where the cores form a bound pair. These results demonstrate new opportunities in producing and manipulating spin states on the nanoscale and may prove useful for new types of ultra-dense storage devices where the information is stored as multiple vortex-core configurations[1,2,3].**

Elementary topological defects of vortex type play an essential role in the general theory of the two-dimensional systems and apply to a variety of phenomena, such as melting, superfluidity, and ferromagnetism[4,5,6]. Spin vortices in ferromagnetic films, which have recently been receiving much attention from the research commuity[1-3,7,8,9,10], have perhaps the greatest variety of behaviors as they possess two types of topological charges – the vortex *chirality*, which is similar to *circulation* in a fluid, and the *polarity* or *polarization* of the vortex core, which is the spin direction out of the plane at the axis of the vortex and has no analogy in fluids. The size of the vortex core is of the order of the exchange length in the material, about 15 nm in Permalloy[11]. It has been recently shown that fast field excitations[1,2] or current pulses[12] can be used to obtain a desired polarity of a vortex core. This demonstration makes spin vortices in submicron ferromagnetic elements a possible candidate for storing information, where the polarization of the core would represent binary "0" and "1". Most of the studies of spin vortices to date have focused on single-layer ferromagnetic film elements containing one vortex[1-3,9,12] or arrays of such ferromagnetic elements each containing one vortex[13,14,15,16]. The inter-vortex interaction is magnetostatic in nature and decays rapidly with the distance away from the individual element, so the core-core interaction is negligible in the array geometry where the individual magnetic elements in the vortex state are spaced apart by lithographic means. The dynamics of two in-plane vortices created in a *single* thin ferromagnetic particle of a few micrometers in size have recently been investigated[17,18,19,20] find that the polarization of the cores affect the sub-GHz gyrational modes of the vortices, even though the direct core-core coupling over their relatively large ~1 µm separation in the plane of the ferromagnetic layer is negligible. The ideal system for exploring the core-core interaction in spin vortex pairs is a ferromagnetic bi-layer with two magnetic layers in the vortex state and separated by a very thin nonmagnetic, much thinner than the vortex core size. Due to the immediate spatial proximity across the thin spacer the two vortex cores can produce bound core-core states. These states with parallel and anti-parallel core polarizations are well separated in energy and therefore should exhibit quite different static and dynamic properties, generally with eigen-modes at much higher frequencies than those for individual vortecies or vortex pairs without directly interacting cores. Such a vertical vortex-pair geometry has recently been discussed in the literature in the limit where the direct core-core coupling is insignificant[21,22,23,24]. Strongly coupled core-core states and their dynamics remain unexplored. In this work we produce parallel/antiparallel core/chirality vortex pairs and investigate their static and dynamic properties experimentally, analytically, and micromagnetically. We find core-core resonances in the 1-10 GHz range, which resemble the rotational and vibrational modes found in the dynamics of bi-atomic molecules.

The samples in this study are nanopillars incorporating an Al-O magnetic tunnel junction between a fixed flux-closed reference layer and a magnetically soft bi-layer, produced by the process described in detail in[25]. This soft bi-layer consists of two Permalloy (Py) layers having about 10 Oe of intrinsic anisotropy, separated by 1 nm thin TaN spacer. The patterned junctions have the in-plane dimensions of 350x420 nm for inducing a small (~100 Oe range) shape anisotropy in the Py layers and the thickness of each Py layer is 5 nm. The spacer material was chosen to make the RKKY coupling across the spacer negligible, so the interaction between the two Py layers is purely dipolar. The reference layer is an exchange-pinned flux-closed CoFeB artificial antiferromagnet producing essentially no fringing field at the soft layer and is used solely for electrical readout. The magnetic tunnel junction resistance is ~1 kOhm and the magnetoresistance is ~20%. The global ground state of the soft bi-layer has the layer moments aligned anti-parallel, forming a flux-closed pair illustrated in the insets to Fig. 1(a). The angle between the average magnetizations in the bottom Py layer and the top reference layer determines the magnetoresistance of the junction. Thus, our built-in read out is magneto-resistive and senses the magnetic configuration of one of the Py layers (bottom), with the state of other (top) Py layer contributing through its effect on the bottom Py layer. The spins in the soft layers can be rotated by application of external fields of intermediate strength (10-100 Oe), whereas the magnetization of the reference layers are rigidly aligned along the long axis of the elliptical particles. The high resistance state corresponds to the anti-parallel state of the tunnel junction, in which the magnetization of the bottom soft layer is aligned opposite to the magnetization of the top reference layer and, vice versa, the low resistance state corresponds to their parallel mutual alignment, as illustrated in Fig. 1(a). The quasi-static field was applied using an electromagnet, while the high frequency fields were applied using an embedded 50 Ohm write line, electrically decoupled from the nanopillar. The AC field was in the plane of the magnetic layers directed at 45 degrees to the long axis of the elliptical particles and had the frequency range of 1 MHz to 10 GHz for continuous wave and down to 80 ps rise/fall time for pulse fields. The quasistatic magnetoresistance was measured by stepwise changing the external magnetic field while recording the resistance of the tunnel junction stack. The dynamic response was measured using the method of microwave spectroscopy, where the resistance was recorded while sweeping the frequency of the ac field excitation, with the variation in resistance proportional to the magnetization oscillation angle, which in turn is greatly enhanced at spin resonant frequencies (for more details see Methods[26]).

The major loop of the quasi-static magneto-resistance measured with the DC field applied along the long axis of the elliptical nanopillar (the easy axis or EA) with the soft layers in the spin-uniform state is shown in Fig. 1(a). Two stable states of low- and high-resistance at zero field correspond to the parallel (P) and anti-parallel (AP) states of the *tunnel junction*. An EA field of 200 Oe overcomes the dipolar repulsion and saturates the soft layers along the field direction. The transition between the AP state and the saturated state of the *soft bi-layer* proceeds through a spin-flop into a scissor state at around ±75 Oe of EA field, followed by a sequence of minor S- and C-type spin configurations, preceding the full saturation at >200 Oe[26]. This major loop behavior is well understood. We recently observed that certain field excitations can drive spin-flop bi-layers into stable resistance states intermediate between high- and low-resistance[27] and suggested that these states are likely due to vortex pairs forming in the soft bi-layer. In this work we are able to identify the individual vortex pair states by measuring and modeling their quasistatic and dynamic behavior and, in particular, analyze the previously unexplored dynamics of spin-vortex pairs with strong core-core interaction.

The net in-plane magnetic moment of the soft layer in the spin vortex state is zero and therefore the resistance of the pillar is precisely intermediate between the high- and low-resistance spin-uniform states. A field applied along the easy axis, which coincides with the easy axis of the reference layer, moves the vortex core along the hard axis and thereby increases the portion of the spins in the soft layer directed along the reference layer. This causes a gradual change in the resistance of the pillar, as shown in Fig. 1(b). This behavior is fully reversible for fields smaller than the vortex annihilation field, ~80 Oe in our case. In the limit of negligible core-core coupling, the shape of the resistance versus field curve should be linear at low fields. Fig. 1(b) clearly shows that the vortex-state *R*(*H*) curve is non-linear, having a step-like or a plateau-like character for the two characteristic examples shown, and can therefore be used to differentiate the individual vortex pair states.

Each vortex has 4 possible states with left- and right-chirality and positive and negative core polarization. Therefore, generally, a vortex pair in a ferromagnetic bi-layer can have the total of 16 configurations. For an ideal bi-layer, the symmetry dictates that 4 sub-classes are different in energy, and each sub-class contains 4 degenerate states obtained by basic symmetry operations. We define the 4 base vortex-pair states as: parallel vortex cores and anti-parallel chirality (P-AP), anti-parallel vortex cores and anti-parallel chirality (AP-AP), parallel vortex cores and parallel chirality (P-P), and anti-parallel vortex cores and parallel chirality (AP-P). These states are illustrated in Fig. 2(a). The relatively strong dipole coupling in the bi-layer translates into a much higher probability to create a vortex pair with the two vortices having AP chirality[27]. This strong preference for the AP-chirality states is further enhanced by the field excitation at the optical spin resonance in the system[26], which drives the Py moments in opposite directions and which we use to generate the vortex states. We note that the AP-chirality states are most interesting as the neighboring vortex cores must move in opposite directions under applied magnetic fields, which is ideal for studying the changes in the behavior of the system while the cores are coupled and decoupled. The P-chirality states, on the other hand, respond to the excitation field by shifting the cores in the same direction, with the core-core coupling unaffected by the excitation – a rather straightforward behavior. In what follows, we therefore concentrate on the AP-chirality vortex states with P and AP vortex core alignment, which represent vortex pairs with strong and weak core-core coupling, respectively.

In order to reliably differential these two P-AP and AP-AP states we perform a detailed micromagnetic modeling of the system of two dipole coupled elliptical particles, first of the non-linear quasi-static $R(H)$ response, such as illustrated in Fig. 1(b). The reference layer of the structure was considered as fully compensated with no stray field, which is a good approximation for our samples. We use the OOMMF micromagnetic package[28] and pay a particular attention to fine meshing the bi-layer structure. We choose the mesh of 1.0x1.0x2.5 nm since we need the nanometer-small vortex cores to be resolved with a high precision to trace their dynamics and avoid mesh pinning which can significantly influence the results. We find that larger mesh sizes used in the literature[22] for modeling spin vortex dynamics do not appropriately model the core-core modes. Typical material parameters for Py were used: saturation magnetization $M_s$=840 x $10^3$ A/m, exchange constant A=1.3 x $10^{-11}$ J/m, damping constant α=0.013. The actual measured intrinsic anisotropy of the Py layers of 5-10 Oe was used. Two complementing and essentially equivalent approaches were used to extract the spin dynamics of the system to be discussed below. Namely, the continuous-wave excitation spectroscopy method which is efficient for modeling fast GHz modes, and the Fourier transform of the magnetization response to a pulse excitation which is efficient for modeling slow sub-GHz gyrational modes. The two methods were found to produce essentially the same results, even though quantitatively the two approaches can be somewhat different. Below we use the results of the pulse-FFT method, since it is vastly more efficient computationally for obtaining full resonance spectra. The micromagnetically simulated $R(H)$ is shown in Fig. 2(b) for three vortex-pair configurations. The response of the two P-chirality states is, as expected, linear in field. The two cores move in the same direction perpendicular to the field irrespective of their mutual alignment. Interestingly, the response of the AP-chirality states is non-linear – step-like at $H$=0 for the AP-AP state and plateau-like for the P-AP state, in good agreement with the experiment of Fig. 1(b). This allows us to uniquely identify the two vortex-pair states observed. We illustrate the mechanism behind this non-linear behavior in more detail in Fig. 3. The applied field favors the spins aligned parallel to the field and disfavors the spins aligned anti-parallel to the field, which leads to an asymmetry in the spin distribution and the vortex core moving toward the disfavored side. If the two vortices have AP chiralities, then their cores must move in opposite directions. They do that if the core-core coupling is small on the scale of the Zeeman energy due to the applied field. The cores of opposite polarization in the AP-AP vortex-pair state have strong on-axis mutual repulsion and a much weaker off-axis attraction. A geometrical consideration suggests that the cores in the 5 nm thick Py layers separated by the 1 nm thin spacer should repel strongly when on-axis, as near-monopoles. The ground state is then an off-axis AP-aligned core pair, with a relatively weak core-core coupling due to the significant separation of the core poles, of the order of the core diameter: ~10 nm spaced dipoles versus 1 nm spaced near-monopoles. The expected difference in the effective core-core coupling is then one to two orders of magnitude. This means that the cores in the AP-AP state will separate easily in the applied field. Fig. 3(a) indeed shows that the cores are well separated already at 15 Oe, with the separation distance roughly proportional to the field strength except at near zero field (~1 Oe) where the core-core axis rotates from being along the hard axis to being along the easy axis, seen as a step in $R(H=0)$. In contrast, the P-AP state has the ground state with the cores *on-axis*, coupled strongly by the near-monopole force across the 1 nm spacer. This strong coupling persists to 20-30 Oe, at which point the cores decouple in a discontinuous fashion, as illustrated in Fig. 3(b). The result is a plateau at low fields and step-like transitions in $R(H)$ at the core-core decoupling field, which is indeed observed in Figs. 1(b), 2(b). The P-chirality states have the expected response where the cores move together in the same direction and the core positions and therefore $R(H)$ are linear in the whole sub-annihilation field range [Fig. 3(c), 2(b)].

Thus, we are able to reproduce and identify the individual vortex-pair states and, significantly, realize the previously unexplored limit where the direct core-core coupling is strong. In this limit, we expect not only changes in the quasi-static behavior but also new spin resonance modes since the energetics of the core movement is modified compared to the case of individual cores or non-interacting cores in a pair. The experimental results below will show that different vortex-pair state (P-AP and AP -AP) exhibit different and rather complex spin-dynamic spectra, with three main resonant modes – a low frequency mode and a weakly-split high-frequency doublet. It is desirable to first gain an insight into the dynamic behavior of the system through an analytical analysis, before interpreting the experimental data and performing numerical micromagnetic modeling. We note that a similar spectral layout, except for the core-core resonance modes, was found in numerical studies of the dynamics of a single vortex in easy-plane two-dimensional ferromagnets[29,30,31] and recently in Permalloy particles[32]. The low-frequency mode corresponds to a gyroscopic motion of an individual core[33] and can be described by the so-called Thiele equation $(\mathbf{G} \times d\mathbf{X}/dt) = \mathbf{F}$, where $\mathbf{X}$ is the vortex core coordinate, $\mathbf{G} = G\mathbf{e}_z$ is the familiar gyrovector with constant $G = 2\pi LM_s/\gamma$, $\mathbf{F} = -\nabla U(\mathbf{X})$ is the force acting on the core, $U(\mathbf{X})$ the potential energy of the core with dissipation neglected. For describing only the low frequency dynamics, it would be sufficient to construct a coupled system of two such equations for two vortices at positions $\mathbf{X}_1 = (x_1, y_1)$ and $\mathbf{X}_2 = (x_2, y_2)$, in upper and lower layers, in which the forces are $\mathbf{F}_{1,2} = -\partial U / \partial \mathbf{X}_{1,2}$, and the energy $U = U(\mathbf{X}_1, \mathbf{X}_2)$ has then two contributions – the interaction of the vortex with the particle boundary and the magnetostatic in nature core-core interaction, described in detail in Supplementary. However, the description of the high-frequency doublet, especially the doublet splitting, is a much more complicated task. The splitting is connected with the topological properties of the core[29,30,31,32,34,35] and the quantitative theory of the doublet for a single vortex with full account for non-local magnetic dipole interactions was constructed only recently[35]. In this regard, an exact analytical solution based on the Landau-Lifshitz equations for two vortices in two different and interacting particles appears to be hopelessly difficult. On the other hand, a phenomenological approach for obtaining the full description of the dynamic spectra was proposed in Ref. 29 and analytically justified for the case of an easy-plane ferromagnet in our Ref. 30. This approach was successfully used for a single vortex and is based on the generalized Thiele equation, which takes into account an inertial term of mass $M$ as well as a non-Newtonian high-order gyroscopic term with third-order time derivative of the core position and with higher-order gyrovector $\mathbf{G}_3 = G_3 \mathbf{e}_z$:

$$(\mathbf{G}_3 \times \frac{d^3 \mathbf{X}}{dt^3}) + M \frac{d^2 \mathbf{X}}{dt^2} + (\mathbf{G} \times \frac{d\mathbf{X}}{dt}) = \mathbf{F}. \quad (1)$$

Recently the validity of this equation for describing the full dynamics of a single vortex motion was verified micromagnetically for thin circular Permalloy particles and explicit expressions for phenomenological constants $G_3$ and $M$ were obtained[36]. We discuss the foundation of the generalized Thiele approach in more detail in Supplementary. Here we only stress that the approach of Refs. 29,30,31 takes into account simultaneously the effective mass $M$ and the higher-order gyrotropic term described by $G_3$. While $G_3$ determines the mid-frequency of the doublet, the mass contributes to the doublet splitting. The big advantage of this phenomenological approach is that, similar to the standard Thiele equation for the low-frequency dynamics of the system, the problem essentially reduces to finding the potential energy. This potential energy has two contributions – the interaction of the vortex with the particle boundary and the magnetostatic in nature core-core interaction, analyzed micromagnetically below and in detail analytically in the Supplementary material:

$$U = \frac{1}{2}\left[k_1(x_1^2 + x_2^2) + k_2(y_1^2 + y_2^2)\right] + U_{\text{Core-Core}}(a), \ a = |\mathbf{X}_1 - \mathbf{X}_2|. \quad (2)$$

For the interaction of the vortex with the particle boundary it is sufficient to consider the linear approximation. Since our ferromagnetic layers are slightly elliptical in shape (aspect ratio 1.2) we introduce separate easy- and hard-axis coefficients to represent the restoring force for the cores. Function $U_{\text{Core-Core}}(a)$ determines the strength of the core-core coupling. The dynamics of the core coordinates are determined by the system of equations of type of Eq. (1) for $\mathbf{X}_1$ and $\mathbf{X}_2$, in which the forces are $\mathbf{F}_{1,2} = -\partial U / \partial \mathbf{X}_{1,2}$. It is easy to show that this system separates into the following two uncoupled sub-systems for variables $x = X_1 - X_2$, $y = Y_1 - Y_2$ and $\bar{x} = X_1 + X_2$, $\bar{y} = Y_1 + Y_2$. A uniform field excites only the first pair of variables, for which the generalized 3-rd order Thiele equations (1) become

$$-G_3 \frac{d^3 y}{dt^3} + M \frac{d^2 x}{dt^2} - G \frac{dy}{dt} + [k_1 + 2\kappa(a)]x = 0,$$
$$G_3 \frac{d^3 x}{dt^3} + M \frac{d^2 y}{dt^2} + G \frac{dx}{dt} + [k_2 + 2\kappa(a)]y = 0, \quad (3)$$

where $\kappa(a) = 2dU_{\text{Core-Core}}(a)/da^2$. These equations fully account for the nonlinear character of $U_{\text{Core-Core}}(a)$, but do not have an exact general solution. We proceed with the limit of interest in this work, that of small oscillations about the equilibrium. We start with the most interesting case of the P-AP configuration, which corresponds to the two cores attracting each other and located in the center of the particles at equilibrium. In the linear approximation the coefficients in (3) become constants, $k_{1,2} + 2\kappa(a) \to k_{1,2} + 2\kappa$. Then, the solutions have the following form,

$$x = \sum_\alpha A_\alpha \cos(\omega_\alpha t + \varphi_\alpha), \quad y = \sum_\alpha B_\alpha \sin(\omega_\alpha t + \varphi_\alpha), \quad (4)$$

where $\varphi_{0,\alpha}$ are the initial phases, index α, α = 0, 1, 2 enumerates three eigen-modes, with the eigen-frequencies given by

$$\omega^2 (G_3 \omega^2 - G)^2 - [M\omega^2 - (k_1 + 2\kappa)][M\omega^2 - (k_2 + 2\kappa)] = 0, \quad (5)$$

and the amplitudes of the eigen-modes $A_\alpha$ and $B_\alpha$ are related by

$$\frac{A}{B} = \sqrt{\frac{k_1 + 2\kappa - M\omega^2}{k_2 + 2\kappa - M\omega^2}}. \quad (6)$$

In the absence of intrinsic anisotropy, this equation for the collective coordinates $x$, $y$ coincides with that for an individual vortex core, except for the coupling parameter $\kappa$, which dominates the potential energy of the core-core interaction in the P-AP state. Using for the elastic constants $k_1$, $k_2$ estimate $k_{1,2} = 20\pi M_s^2 L^2 / 9R_{1,2}$, where $R_{1,2}$ is the maximum and minimum radii of the sample and $\kappa$ for the geometry in question (very low aspect ratio) shows that the effective anisotropy of the core motion is negligible and, therefore, $A_\alpha = B_\alpha$ and $\bar{k} = \sqrt{k_1 k_2} \approx k = 40\pi M_s^2 L^2 / 9R$ (see Ref. 33). The obtained eigen-modes can be classified as a lower-frequency mode with frequency $\omega_0^{P-AP}$, and a higher-frequency weakly-split doublet with frequencies $\omega_1 = \bar{\omega} - \Delta\omega$, $\omega_2 = -(\bar{\omega} + \Delta\omega)$. In the limit $\omega_0^{P-AP} \ll \bar{\omega}$, $\omega_0^{P-AP} = (\bar{k} + 2\kappa)/G$, and the exact values of the eigen frequencies are determined by Eq. (5). Estimating the eigen-frequencies for the P-AP state yields $\omega_0/2\pi$=2.6 GHz, which compares well with the measured value of ~2 GHz (see below). As to the high-frequency doublet, good agreement with the experiment is obtained using numerical coefficients $\eta_M \approx 0.58$ from the single-vortex case (Ref. 36) and $\eta_{G3} = 0.312$, which is a factor of two smaller than the corresponding single-vortex coefficient [Supplementary]. The resulting $G_3$ and $M$ yield $\omega_1 = 5.26$ GHz and $|\omega_2| = 6.49$ GHz, which agree well with the measured eigen-frequencies (2, 5.4, 6.3 GHz, as shown below).

Next, let us consider the AP-AP case, in which the equilibrium configuration corresponds to the vortex cores shifted off-center, along the longer axis of the particles: $X_1 = a_0/2, X_2 = -a_0/2$, with $a_0 = 18.8$ nm for our material system. For this AP-AP configuration we arrive at the coupled system of equations similar to (3), however, the dynamics of the system is quite different and, in particular, the anisotropy plays a significant role. Our analysis of the AP-AP state results in a modification of Eq. (5) with the $k_1 + 2\kappa(a)$ term replaced by $\kappa_\parallel$, and with the coefficient of the restoring force along the $y$ axis determined solely by the anisotropic interaction of the vortex with the particle boundary. Specifically, $k_2 + 2\kappa(a) \to (k_1 - k_2)$ in Eq. (5). The frequency of the gyroscopic mode is then easily found as $\omega_0^{AP-AP} = \frac{1}{G}\sqrt{2\kappa_\parallel(k_1 - k_2)}$. Since difference $k_1 - k_2$ is small, $\omega_0^{AP-AP}$ is much lower than the corresponding $\omega_0^{P-AP}$, which is in agreement with the experiment. Another significant difference between this resonant core-core mode in the AP-AP state as compared to the P-AP state is that the core motion proceeds along a highly elongated ellipse, with the axis ration given by $B/A = \sqrt{(k_1 - k_2)/2\kappa_\parallel} \ll 1$.

The anisotropy of $k$ is not known exactly and its analysis goes beyond the scope of this paper. For an estimate we use the value obtained by substituting $R$ with $R_1$ or $R_2$ for $k_1$ or $k_2$, respectively. This yields $\omega_{AP-AP} \approx 0.2$ GHz, which is in qualitative agreement with the observed 0.34 GHz (see below). The difference is not surprising, since the above rough estimate underestimates the actual anisotropy and therefore lowers the estimated $\omega_0^{AP-AP}$. On the other hand, analysis shows that the value of $k_1 - k_2$ practically does not influence the high-frequency eigen-modes. Thus, taking the same $\eta_M$ and $\eta_{G3}$ as for the P-AP state, the AP-AP high-frequency doublet is expected at ~5.8 GHz and ~6.5 GHz, which again is in good agreement with the measured doublet frequencies of 5.9 GHz and 6.4 GHz.

Our analysis thus shows that the dynamics of the two types of bound core-core pairs differ in a principle way. Most notably the character of the lower-frequency mode changes on changing the core-core coupling from P to AP – the frequency is lower by an order of magnitude and the trajectory becomes highly elliptical from being almost circular. It is remarkable that the values of all six eigen-frequencies are well described by a simple equation of third order, with the same values of the two phenomenological coefficients, $M$ and $G_3$, for P and AP coupled vortex cores. It is interesting to note that the value of the effective mass $M$ is the same and that of $G_3$ twice smaller compared to the values found for individual spin vortices. This difference in $G_3$ is evidence of the non-locality of this dynamic characteristic (see Supplementary).

The above analysis allows to uniquely identify the nature of the three eigen-modes. The ~2 GHz mode in the P-AP case represents a fixed-orbit gyration or rather a rotation of the cores about the center-of-mass of the core-core pair, symmetrically displaced from the origin by the excitation field. In the AP-AP case this mode is found much lower in frequency and the core orbits are highly elliptical. The high-frequency doublet represents a smaller-amplitude precession about this rotational orbit, where the cores precess in opposite directions ($\omega_1$ and $\omega_2$ have opposite signs). In the absence of the core-core coupling, the potential force acting on the individual cores is greatly reduced, which results in an order of magnitude lower gyrational frequency ($\omega_0$ ~ 0.1 GHz) and a modified high-frequency doublet. Thus, the spin dynamics of the vortex pair crucially depends on whether the vortex cores are coupled or decoupled, with the potential energy of the core-core interaction shifting the main rotational resonance by a factor of 10 in frequency.

Microwave spectroscopy data for the most interesting case of the P-AP state, where the cores are strongly coupled at zero field and can be decoupled by applying a dc field of 30-40 Oe, are shown in Fig. 4(a). The zero-field spectrum has a pronounced peak at slightly above 2 GHz and a well-defined double peak at ~6 GHz. The relatively broad 6-7 GHz peak of the doublet appears to have a weakly defined sub-structure. Interestingly, the sub-GHz region of the spectrum, where the single-core gyrational mode would be expected, is at the noise floor of the measurement and shows no traces of a resonance. When the cores are decoupled by the 40 Oe field applied along the EA (grey line in Fig. 4a), the strong rotational peak as well as the vibrational doublet vanish and the known[33] single-core gyrational mode appears at ~0.2 GHz. This means that the strong on-axis coupling of the two cores, which makes them a bound pair in the P-AP state, suppresses the individual core gyration, such that the lowest-frequency spin excitation becomes the new rotational resonance at 10 times higher frequency, in which the cores precess about their magnetic center-of-mass.

The magneto-resistance spectrum for the same junction set into the AP-AP vortex state is shown in Fig. 4(b). The spectrum recorded at 40 Oe EA field (grey line) is essentially identical to that in the P-AP state at 40 Oe, which indicates that when the cores are non-interacting the dynamics of the vortex pair does not depend on the core polarization. Importantly, the rotational peak at ~2 GHz is absent in the AP-AP state, which reinforces our interpretation of this resonance as due to the strong core-core interaction in the P-AP state [black line in Fig. 4(a)]. The zero-field AP-AP spectrum of Fig. 4(b) shows a pronounced low-frequency peak at ~0.2 GHz, which can be associated with the core-core elliptical rotation and expected to almost overlap in frequency with the individual core gyrational modes. Fig. 4(b) also shows that the low-frequency resonance is accompanied by high-frequency core vibrations seen as a doublet at ~6 GHz. We find that these vibrational modes are effectively suppressed at high fields [40 Oe data in Figs. 4(a,b)], where the cores are significantly displaced from the center of the Py particles.

The measured spectra are therefore in good agreement with the analytically predicted behavior. It is informative to note here that the success of our phenomenological theory is not due to the fact that $G_3$ and $M$ are adjustable parameters. Importantly, the same values of these parameters describe well the two sets of three frequencies, observed for two substantially different experimental configurations (in terms of the dominant interaction strength), P-AP and AP-AP.

The above results are in also good agreement with our micromagnetic simulations shown in Fig. 4(c) for the P-AP state with coupled and decoupled vortex cores. In order to perform the simulation in a reasonable time frame a pulse excitation with a subsequent FFT of the magnetization response ($\mathbf{M}_x$) was used instead of the continuous-wave (CW) method. This approach yields relatively lower amplitudes of the highest frequency resonance modes as they are not efficiently excited by the pulse excitation (compared to CW). To better visualize all the simulated modes we show the results on the log scale. Nevertheless some quantitative differences, the simulation reproduces excellently the major change in the spin dynamics of the system from the rotational to the gyrational resonance as the two cores are decoupled, as well as the values of the characteristic gyrational, rotational, and vibrational resonance frequencies.

Fig. 5 illustrates the spin dynamics in the system, focusing on the P-AP state, which exhibits all three resonance modes, pronounced and well separated in frequency. The rotational resonance is illustrated in Fig. 5(a) and consists of a mutual rotation of the two coupled cores. The inter-core separation $d$ is constant for a continuous wave excitation of fixed amplitude, or decaying in time if the excitation is a fast pulse as illustrated in Fig. 5(a-c). The trajectories of the two cores (black and red lines) are circular and out of phase, with the radius of ~1 nm. An interesting analogy to this core-core resonance is the rotational resonance in a diatomic ionic molecule: the top and bottom magnetic monopoles of the core-surfaces facing each other play the role of the ionic charges and the magnetostatic coupling between the monopoles substitutes the Coulomb attraction in the molecular case. Superposed on this nearly circular trajectory are core vibrations with the typical amplitude of ~0.1 nm, as shown in Fig. 5(b), in which the cores vibrate out of phase about the circular rotational orbit. When the cores are decoupled at 40 Oe, they gyrate independently, with the typical orbit size of ~10 nm, as illustrated in Fig. 5(c). Thus, the micromagnetic model provides an accurate and illustrative mapping of the core dynamics for all three main resonance modes in the system, and agrees well with the experimental and analytical results.

In conclusion, the dynamics of vortex pairs in ferromagnetic bi-layers are investigated with an emphasis on the core-core interaction in the system. We show how individual vortex-pair states, having different combinations of the vortex chirality and core-polarization, can be identified using their response to static and dynamic fields, which potentially can be used for storing bits of information as vortex core/polarity states in magnetic nanostructures. The observed spin dynamics is explained in terms of collective core-core rotational and vibrational modes, which, interestingly, mimic the rotational and vibrational resonances in diatomic molecules. We generalized the Thiele theory for the dynamics of strongly interacting vortex core pairs and successfully verify it on the experiment.

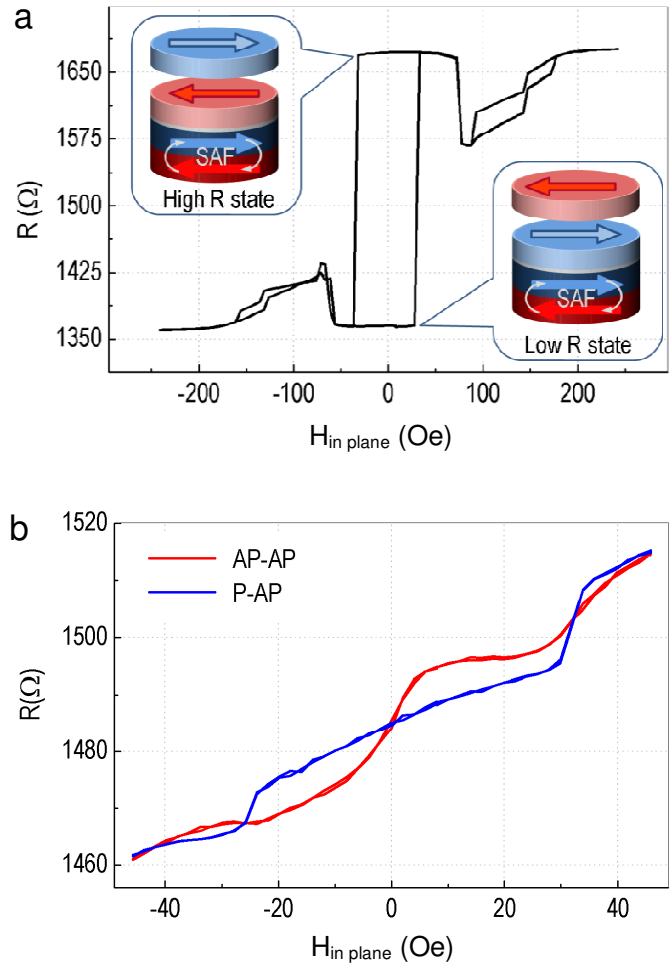

**Fig. 1:** (a) Major magneto-resistance loop of a spin-flop tunnel junction, 350x420 nm in size, showing two antiparallel ground states of the soft bi-layer corresponding to low- and high-resistance states of the junction. (b) Minor magneto-resistance loop for the same junction set in a vortex magnetization state with the resistance intermediate between the low-*R* and high-*R* states shown in (a). Two typical vortex states are shown, with *R* vs *H* step-like and plateau-like at zero field. The vortex annihilation field is approximately 80 Oe.

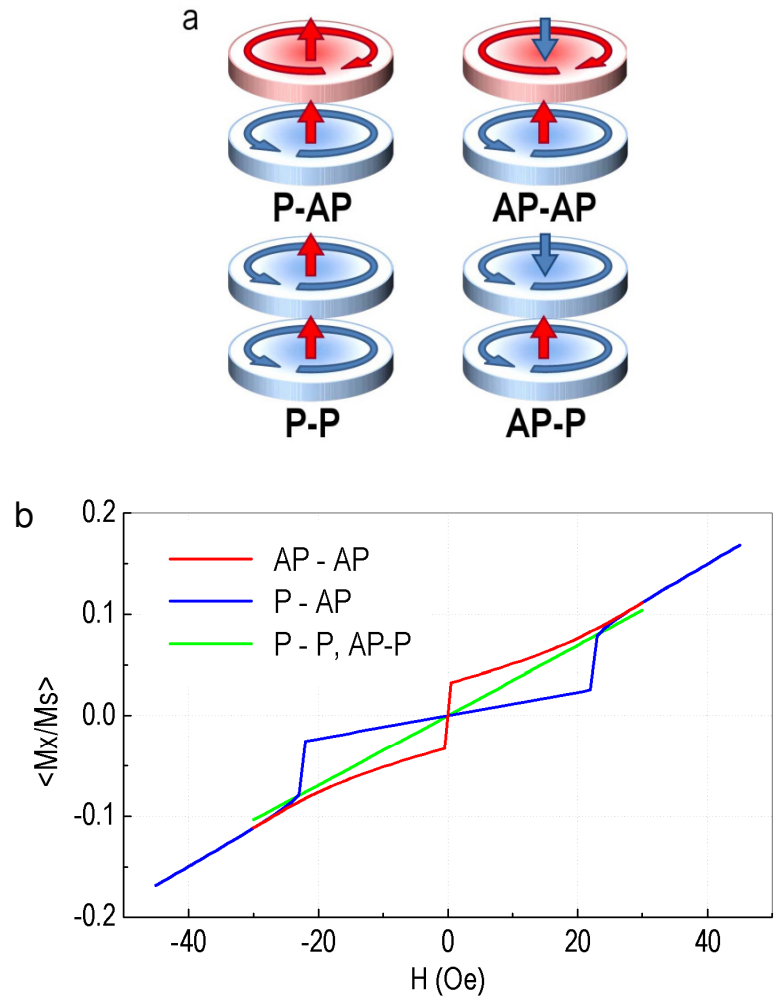

**Fig. 2:** (a) 4 non-degenerate in energy vortex-pair states in a ferromagnetic bi-layer: parallel core alignment and antiparallel chirality, P-AP; antiparallel cores and antiparallel chiralities, AP-AP; parallel cores and parallel chiralities, P-P; antiparallel cores and parallel chiralities, AP-P. (b) Micromagnetic simulation of the relative magneto-resistance ($<M_x/M_S>\sim R$) for the vortex pair states in 350x420 nm elliptical dipole coupled bi-layer shown in (a).

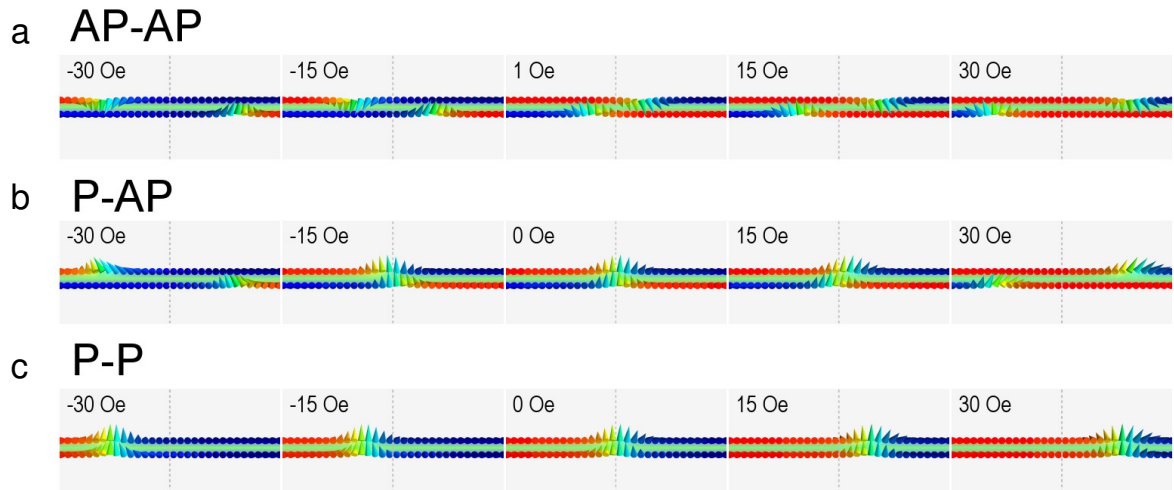

**Fig. 3:** Hard axis cross sections of the micromagnetically simulated spin distribution for three vortex-pair states and three characteristic applied field strengths. The applied field is along the easy axis and causes the vortex cores to move along the hard axis, perpendicular to the applied field. The core are weakly coupled in the AP-AP state and readily separate in 15 Oe. In contrast, approximately 30 Oe is needed for decoupling the core-core pair in the P-AP state, while the field of 15 Oe produces only a small displacement of the pair, ~1 nm. The cores in the P-chirality states (P-P state shown) move in the same direction and never decouple.

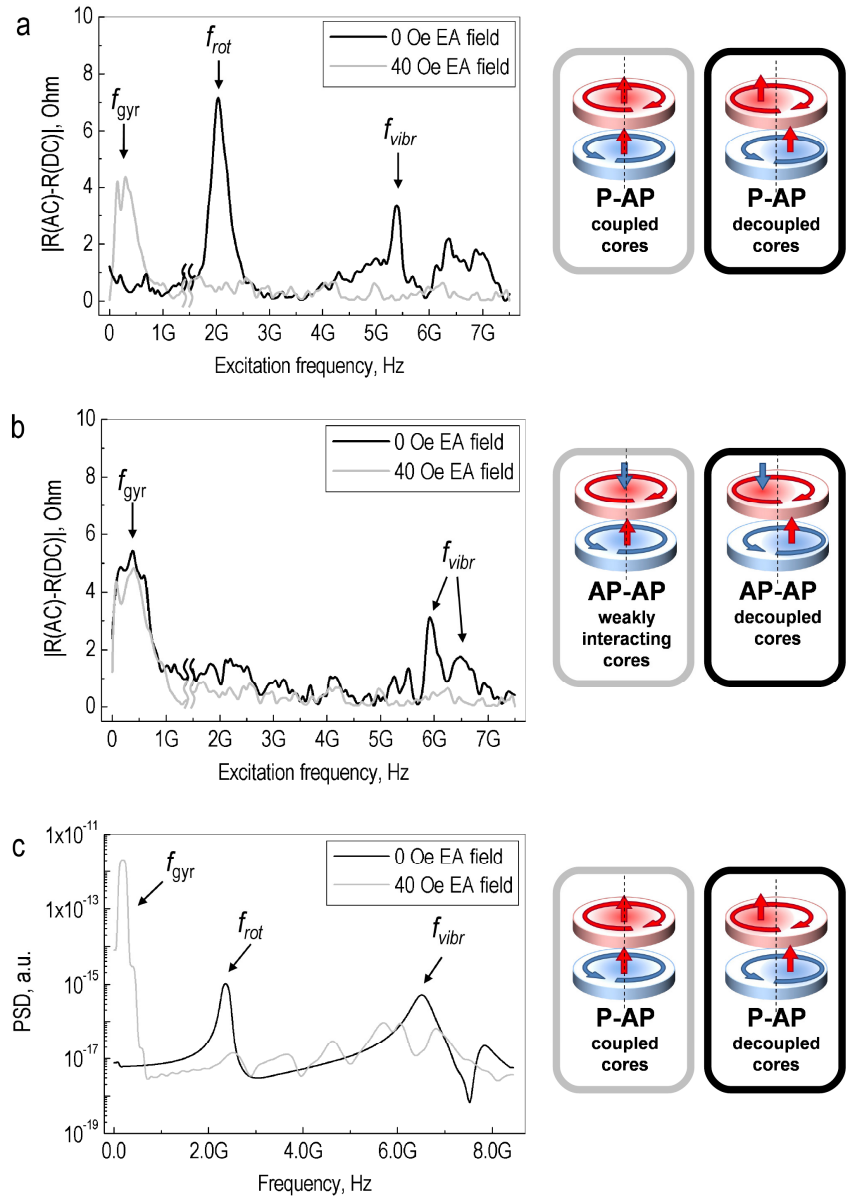

**Fig. 4:** Microwave spectroscopy data for (a) P-AP vortex-pair state showing ~2 GHz and ~6 GHz rotational ($f_{rot}$) and vibrational ($f_{vibr}$) resonances at zero field, when the cores are strongly coupled (black). Application of 40 Oe decouples the cores and the spectrum transforms into a single-core gyration (grey) mode ($f_{gyr}$). (b) Spectra for the same junction set into the AP-AP state. The weakly interacting cores gyrate essentially independently. The vibrational doublet is visible at zero field. In (a) and (b) the AC amplitude was 7.5 Oe above 1.5 GHz and was reduced to 5 Oe at below 1.5 GHz to avoid annihilating the vortex states at over-critical core velocities[37]. (c) Micromagnetically simulated P-AP spectra corresponding to measured spectra in (a).

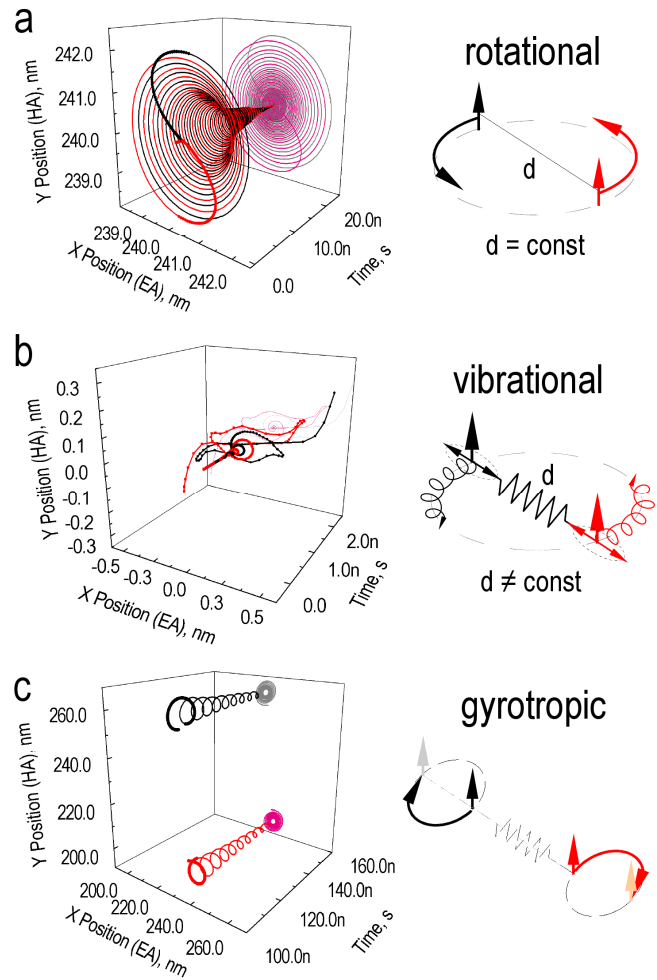

**Fig. 5:** Micromagnetic illustration of the vortex core trajectories for the three resonance modes in the P-AP state. The magnetization response is to a 100 ps field pulse, shown versus time as the oscillation amplitudes vanish and the core positions decay to their steady state values (particle center). The back plane shows the *x-y* projection of the core motion. At zero field, the strong core-core coupling suppresses the single-core gyrational mode and yields a collective rotational mode at ~2 GHz with the typical radius of ~1 nm, shown in (a). Slight (~0.1 nm radius) out of phase vibrations at ~ 6 GHz are superposed on the rotational orbit, as shown in (b). The cores decoupled by a 40 Oe field gyrate independently at ~0.2 GHz in orbits of ~10 nm diameter, as can be seen in (c).

# Supplementary material

## Core-core coupling

The sources of the field mediating the core-core interaction are the magnetic poles on the top and bottom surfaces of the two magnetic layers, which have the strength $M_s \cos\theta(x,y)$ and $-M_s \cos\theta(x,y)$, where angle $\theta(x,y)$ determines the out of plane component of the core magnetization, exponentially localized in the vicinity of the core. The interaction energy for two vortices includes the interaction of such magnetic poles on four surfaces, separated by distances $D$, $2L+D$, and (twice) $L+D$, as shown in Fig. S1.

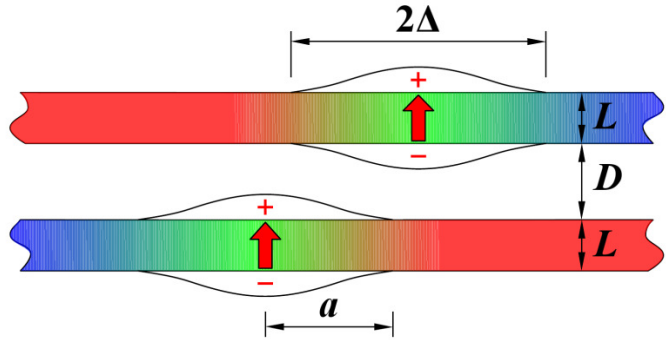

**Figure S1**. Schematic of the core-core coupling model. Red arrows indicate the positions and polarizations of the vortex cores. Dimensions $D$ and $L$ are not to scale.

The total potential energy of the interaction is given by the following simple expression

$$U_{core-core} = \sigma M_s^2 \left[ 2F(L+D) - F(D) - F(2L+D) \right],$$

where $\sigma = \pm 1$ describe the relative core polarizations, $\sigma = -1$ and $\sigma = 1$ correspond to the P-AP and AP-AP cases, respectively, and function $F(X)$ is given by

$$F(X) = \int \frac{\cos\theta_0(\mathbf{r}_1 + \mathbf{a}/2) \cos\theta_0(\mathbf{r}_2 - \mathbf{a}/2)}{\sqrt{(\mathbf{r}_1 - \mathbf{r}_2)^2 + X^2}} d\mathbf{r}_1 d\mathbf{r}_2 .$$

Here $\mathbf{r}_1$ and $\mathbf{r}_2$ are two-dimensional vectors corresponding to the first and second layers, $\cos\theta_0(\mathbf{r})$ is the standard distribution of the out-of plane magnetization with the vortex core placed at the origin, $\cos\theta_0(0) = p = \pm 1$, $\mathbf{a}$ describes the relative displacement of the vortices (in-plane separation of the cores). We use the well-known Feldkeller-Thomas approximation, in which the spin distribution within the core $\theta(x, y)$ is taken to be Gaussian, $\cos\theta(\mathbf{r}) = \exp(-r^2/\Delta^2)$. The value of $\Delta$ is determined by the condition that the integral $M_s \int \cos\theta(\mathbf{r}) d\mathbf{r}$ coincides with that obtained using exact numerical micromagnetic calculations of $\theta(x, y)$, which equals $2\pi\xi M_s l_0^2$, $\xi \approx 1.361$, with $l_0 = \sqrt{A/4\pi M_s^2}$ being the exchange length. For Permalloy, the characteristic lengths are found to be $l_0 \approx 5$ nm and $\Delta \approx 8.25$ nm. Using these values, a straightforward calculation yields a simple expression for the potential energy of the core-core interaction in the form

$$U_{\text{Core-Core}} = \sigma M_s^2 \Delta^3 f(u), \quad f(u) = \left[ -\Phi\left(u, \frac{D}{\Delta}\right) + 2\Phi\left(u, \frac{L+D}{\Delta}\right) - \Phi\left(u, \frac{2L+D}{\Delta}\right) \right],$$

where $\sigma = 1$ for the P-AP state, $\sigma = -1$ for the AP-AP state, $u = a/\Delta$, and universal function $f = f(u, \delta)$ is given by the integral

$$\Phi(u, \delta) = \pi^2 \sqrt{2} e^{-u^2/2} \int_0^\infty \frac{r dr}{\sqrt{r^2 + \delta^2/2}} e^{-r^2} I_0(u\sqrt{2}r),$$

which can be easily calculated numerically. The asymptotic behavior of the universal function $f$ at small distances, $a/\Delta \ll 1$, can be represented by a combination of error functions, $f(u) = f(0) + \frac{\alpha}{2} u^2$, where for our geometry $f(0) = -1.6825$, $\alpha = 1.8145$. Our analysis shows that the direct core-core coupling is localized to distances of the order of a few $\Delta$, or $a \leq$ 20-30 nm for Permalloy. For small distances it dominates the vortex repulsion from the sample boundaries. For larger distances $a \gg \Delta, L$ the cores interact as magnetic dipoles, with $U_{core-core}(a) = 4.046 \sigma M_s^2 \Delta^6 / a^3$. These short- and long-range behaviors of the core-core interaction energy are shown in Fig. S2 as dashed lines.

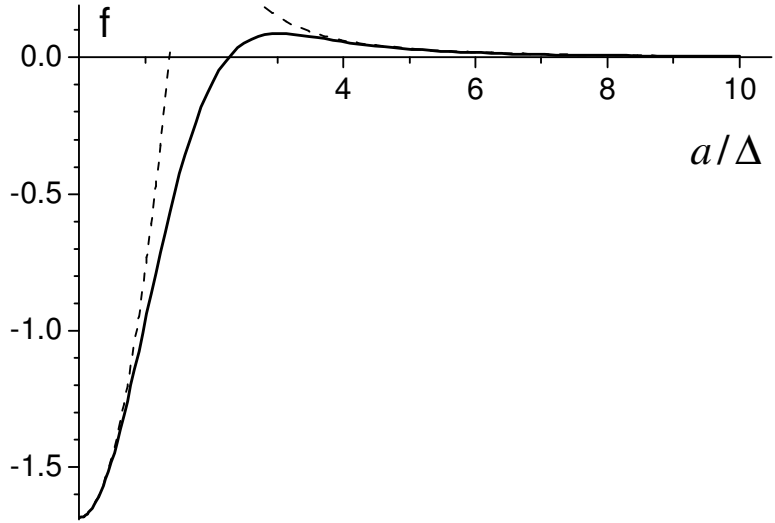

**Figure S2**. The core-core potential in units of $M_s^2 \Delta^3$ for the case of $\sigma = 1$ (P-AP), which corresponds to core-core attraction at small separation distance $a$; the aforementioned short- and long-range asymptotic behaviors are shown by dashed lines.

For the P-AP geometry, both the core-core coupling and the interaction of the cores with the sample boundaries favor the configuration with the vortex cores in the center of the particles, as clearly seen from the energetics of the system shown in Fig. S2. For small deviations of the core positions from the system center, $\delta \mathbf{r}_1$ and $\delta \mathbf{r}_2$, $|\delta \mathbf{r}_{1,2}| \ll \Delta_0$, with $\delta \mathbf{r}_1 = -\delta \mathbf{r}_2$ and $|\delta \mathbf{r}_{1,2}| = a/2$ in our case, the key coupling parameter in Eq. (3) becomes $\kappa = \kappa_{eff}^{P-AP} = 1.8145 M_s^2 \Delta$. Its magnitude is much greater than that of the coefficient of the restoring force, which for our vortex pair is $k = 40\pi M_s^2 L^2 / 9R$. These values are used in Eqs. (3-6) of the main text.

For the AP-AP case, the core-core interaction is strongly repulsive with the cores in the center, and becomes moderately attractive at $a_0 \approx 18.8$ nm for our system (see Fig. S3). The small-signal expansion of the total potential near $a_0$ yields $U \approx U_0 + \kappa_\parallel (a - a_0)^2 / 2$, with the coefficient $\kappa_\parallel$ of order of $\kappa$, $\kappa_\parallel = 0.38\kappa$.

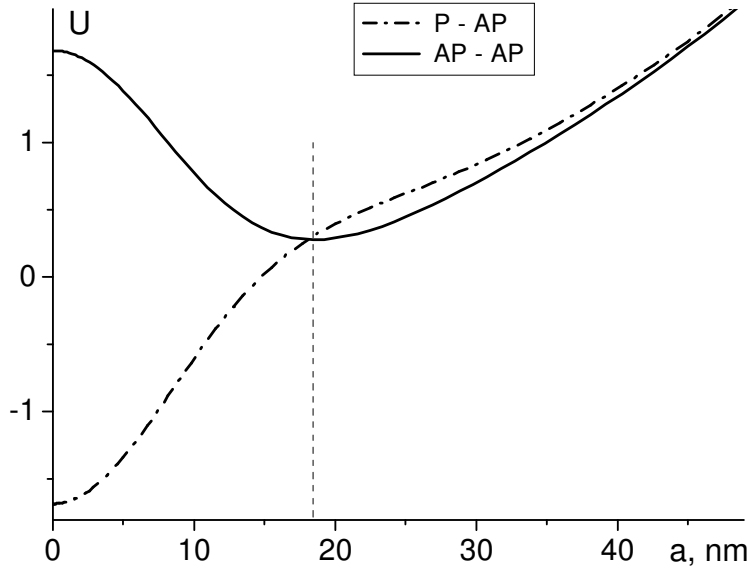

**Figure S3**. The total potential energy of interaction (in units of $M_s^2 \Delta^3$) for two vortices with antiparallel (solid line) and parallel cores (dash-dotted line) separated by distance $a$ near the center of the system. Here we have used the known value of $k$ for circular particles [31] and the mean particle radius for our samples ($\bar{R} = 192$ nm).

## Effective mass and non-locality of third-order gyroscopic term

As regards the choice of coefficient $G_3$, we point out the following. In principle, Eq. (1) is phenomenological and was introduced for analyzing the results of a number of numerical micromagnetic investigations on a variety of magnetic nanostructures, rather than derived from the relevant microscopic considerations. Recently, however, the validity of this equation for describing the full dynamics of a single vortex motion was verified micromagnetically for thin circular permalloy particles and explicit expressions for the phenomenological constants $G_3$ and $M$ were obtained [36]:

$$G_3 = \eta_{G3} \frac{R}{4\pi\gamma^3 M_s}, \quad M = \eta_M \frac{L}{\gamma^2}. \tag{2}$$

Here $\eta_{G3}$ and $\eta_M$ are numerical coefficients of order unity. Their values, $\eta_{G3} \approx 0.63$ and $\eta_M \approx 0.58$, agree well with the analytical theory of the magnon resonance in the system with a single vortex [36].

It was additionally found that the $G_3$-term is nonlocal in nature. In contrast to the local terms scaled by $M$ and $G$, the $G_3$-term does not depend on the vortex length (film thickness), but rather on the sample radius. Thus, increasing the film thickness by, for example, two times does not change $G_3$, while the other coefficients increase twice in value. It is then natural to expect that a strongly coupled vortex pair is characterized by the same value of $G_3$ as a single vortex, and therefore the expected effective value per core is then $G_3/2$. A more detailed analysis of $G_3$ requires a separate investigation and goes beyond the scope of this paper.